\documentstyle [12pt]{article}
\begin{document}
\def\o{\over}
\def\Ar{\rightarrow}
\def\bar{\overline}
\def\r{\gamma}
\def\d{\delta}
\def\a{\alpha}
\def\b{\beta}
\def\n{\nu}
\def\m{\mu}
\def\k{\kappa}
\def\e{\epsilon}
\def\p{\pi}
\def\th{\theta}
\def\om{\omega}
\def\vp{{\varphi}}
\def\Re{{\rm Re}}
\def\Im{{\rm Im}}
\def\ra{\rightarrow}
\def\t{\tilde}
\def\bar{\overline}
\def\l{\lambda}
\def\G{{\rm GeV}}
\def\M{{\rm MeV}}
\def\eV{{\rm eV}}
\baselineskip=24.5pt
\setcounter{page}{1}
\thispagestyle{empty}
\topskip 2.5  cm
%\topskip 0.5  cm
%\begin{flushright}
%\begin{tabular}{c c}
%& {\normalsize hep-ph/9904338}\\
%& April 1999
%\end{tabular}
%\end{flushright}
\topskip 2 cm
\centerline{\large\bf Democratic Mass Matrices} 
\centerline{\large\bf from Broken $O(3)_L\times O(3)_R$ Flavor Symmetry }
\vskip 1 cm
\centerline{M. Tanimoto $^{a,}$
\footnote{E-mail address: tanimoto@edserv.ed.ehime-u.ac.jp},
 T. Watari $^b$ and T. Yanagida $^{b,\ c}$}
\vskip 1 cm
\centerline{$^a$ Faculty of Education, Ehime University, Matsuyama
790-8577, Japan}
\centerline{$^b$ Department of Physics, University of Tokyo, Tokyo
113-0033, Japan}
\centerline{$^c$ Research Center for the Early Universe, University of Tokyo,
Tokyo 113-0033, Japan}
\vskip 1 cm

\vskip 2 cm
\noindent
{\large\bf Abstract}

 We impose $O(3)_L\times O(3)_R$ flavor symmetry in the supersymmetric
standard model. 
  Three lepton doublets $\ell_i$
	transform as an $O(3)_L$ triplet and three charged leptons $\bar e_i$
	as an $O(3)_R$ triplet, while Higgs doublets $H$ and $\bar H$ are 
	$O(3)_L\times O(3)_R$ singlets. 
 We discuss a flavor $O(3)_L\times O(3)_R$ breaking mechanism that leads to
    "successful" phenomenological mass matrices,
    so-called "democratic" ones, in which the large
    $\n_\mu-\n_\tau$ mixing is naturally obtained.
	Three neutrinos have nearly degenerate masses of order $0.1\eV$ 
	which may be accesible to future double $\b$-decay experiments.
 We extend our approach to the quark sector and show that it is well
consistent with the observed  quark mass hierarchies and the CKM matrix
elements.
 However, the large mass of the top quark requires a relatively large
coupling constant.
\newpage
\topskip 0.0 cm
 
 %%%%%%%%%%%%%%%%%%%%%%%%%%%%%%%%%%%%%%
 Yukawa coupling matrices of Higgs field 
 (i.e. masses and mixings of quarks and leptons)
are the least understood part of the standard electroweak gauge theory,
which are, however, expected to be an important hint of more fundamental theory.
There have been a number of attempts to underatand  mass matrices of quarks
and leptons
 by postulating some broken flavor symmetries based on Abelian \cite{Abe} or
 non-Abelinan \cite{NonAbe} groups.  The $O(3)$ flavor symmetry \cite{O31,O32}
 has a unique prediction, that is almost degenerate neutrino masses.
 Using the result of the atmospheric neutrino oscillation observed in
 SuperKamiokande experiments \cite{SKam}, one may conclude
  that all three neutrinos have masses of order $0.1 -1 \eV$ for the case of
   degenerate neutrinos. 
   This is very interesting since such degenerate neutrino masses lie 
   in the region accesible to future double-$\b$ decay experiments \cite{Double}
   if the neutrinos are Majorana particles.
   
   On the contrary, masses of quarks and charged leptons vanish
   in the $O(3)$ symmetric limit.
   Therefore, mass matrices of quarks and leptons are determined by details of
   breaking pattern of the flavor symmetry.  In this letter,
   we discuss a possible flavor $O(3)$ breaking mechanism that leads to
   "successful" phenomenological mass matrices,
   so-called "democratic" ones \cite{Demo1,Demo2}, in which the large
   $\n_\mu-\n_\tau$ mixing suggested from the atmospheric neutrino oscillation
   \cite{SKam} is naturally obtained.
   
   We consider the supersymmetric standard model and impose
  $O(3)_L\times O(3)_R$ flavor symmetry.  Three lepton doublets $\ell_i(i=1-3)$
transform as an $O(3)_L$ triplet and three charged leptons $\bar e_i(i=1-3)$
	as an $O(3)_R$ triplet, while Higgs doublets $H$ and $\bar H$ are 
 $O(3)_L\times O(3)_R$ singlets.  We will discuss the quark sector later.
	
	%%%%%%%%%%%%%%%%%%%%%%%%
	%%%%%%  Breaking %%%%%%%
	%%%%%%%%%%%%%%%%%%%%%%%%
	 We introduce, to break the flavor symmetry, pair of fields
	 $\Sigma^{(i)}_{L} (i=1,2)$ and  $\Sigma^{(i)}_{R} (i=1,2)$ 
	 which transform as symmetric traceless tensor {\bf 5}'s of  $O(3)_L$
	 and  $O(3)_R$, respectively.
	 We assume that the $\Sigma^{(i)}_{L}({\bf 5},{\bf 1})$ and 
	  $\Sigma^{(i)}_{R}({\bf 1},{\bf 5})$ take values
%%%%%%%%%%%%%%%%%%%%%%%%%%%%%%
 \footnote{The more general case for $\Sigma^{(i)}_{L}$ and $\Sigma^{(i)}_{R}$
	  will be discussed in the end of this letter.}
%%%%%%%%%%%%%%%%%%%%%%%%%%%%%%
	\begin{equation}
\Sigma^{(1)}_{L,R} =
\left( \matrix{1 & 0 & 0 \cr
        0 & 1 & 0 \cr  0 & 0 & -2  \cr  } \right )  w^{(1)}_{L,R} \ ,
	\label{S1}
\end{equation}		
\noindent
and 					 
\begin{equation}
\Sigma^{(2)}_{L,R} =
\left( \matrix{1 & 0 & 0 \cr
        0 & -1 & 0 \cr  0 & 0 & 0  \cr  } \right )  w^{(2)}_{L,R}\ .
\label{S2}
\end{equation}		
We consider that these are explicit breakings  of
	$O(3)_L\times O(3)_R$  rather than vacuum-expectation values of  
	$\Sigma^{(i)}_{L,R}$(spontaneous breaking),
	otherwise we have unwanted massless Nambu-Goldstone multiplets.
	In the following discussion we use dimentionless breaking parameters 
	$\sigma^{(i)}_L$ and $\sigma^{(i)}_R$, which are defined as
\begin{equation}
 \sigma^{(1)}_{L,R}\equiv \frac{\Sigma^{(1)}_{L,R}}{M_f} =
  \left( \matrix{1 & 0 & 0 \cr
        0 & 1 & 0 \cr  0 & 0 & -2  \cr  } \right )  \delta_{L,R} \ ,
\end{equation}	
\noindent  and 		
 \begin{equation}
  \sigma^{(2)}_{L,R}\equiv \frac{\Sigma^{(2)}_{L,R}}{M_f} =
  \left( \matrix{1 & 0 & 0 \cr
        0 & -1 & 0 \cr  0 & 0 & 0 \cr  } \right )  \e_{L,R} \ .
 \end{equation}	
 \noindent Here, $M_f$ is the large flavor mass scale,
 $\delta_{L,R}=w_{L,R}^{(1)}/M_f$ and $\e_{L,R}=w_{L,R}^{(2)}/M_f$.
 We assume $\delta_{L,R}, \e_{L,R} \leq 1$.
  
   The neutrinos acquire small Majorana masses from a superpotential,
%%%%%%%%%%%%%%%%%%%%%%%%%%%%%%%%%%%%%%%%%%%%%%%%%%%%%%%%%%%%%%
 \footnote{This superpotential is induced by integrating some massive 
 heavy fields
   $N_i(i=1-3)$ which transform as a triplet of $O(3)_L$(rather than $O(3)_R$).
	In this case the mass $M$ corresponds to  Majorana masses of $N_i$.}
%%%%%%%%%%%%%%%%%%%%%%%%%%%%%%%%%%%%%%%%%%%%%%%%%%%%%%%
\begin{equation}
  W =\frac{H^2}{M}\ell ( {\bf 1}+\a_{(i)}\sigma^{(i)}_L ) \ell  \ ,
\end{equation}	
\noindent
which yields a neutrino mass matrix as
\begin{equation}
  \widehat m_\n =
  \frac{<H>^2}{M}\left \{ \left (\matrix{1 & 0 & 0 \cr
        0 & 1 & 0 \cr  0 & 0 & 1 \cr  }  \right ) + 
		 \a_{(1)} \left (\matrix{1 & 0 & 0 \cr
        0 & 1 & 0 \cr  0 & 0 & -2 \cr  }   \right ) \delta_L +
		\a_{(2)} \left (\matrix{1 & 0 & 0 \cr
        0 & -1 & 0 \cr  0 & 0 & 0 \cr  } \right ) \e_L \right \} .
		\label{nuemass0}
 \end{equation}
  \noindent Here, $\a_{(i)}$ are ${\cal O}(1)$ parameters and the mass  $M$
denotes
  a cut-off scale of the present model which may be different from the
flavor scale $M_f$.
  We take $M\simeq 10^{14-15} \G$ to obtain 
  $m_{\n_i}\simeq 0.1- 1 \eV$ indicated from the atmospheric neutrino
oscillation
  \cite{SKam} for degenerate neutrinos.
  
  The above breaking is, however, incomplete, since the charged leptons
  remain massless. We introduce an $O(3)_L$-triplet and an
   $O(3)_R$-triplet fields
   $\phi_L({\bf 3},{\bf 1})$ and $\phi_R({\bf 1},{\bf 3})$ to produce masses of
   the charged leptons.  The vacuum expectation values of 
   $\phi_L$ and $\phi_R$ are determined by the following superpotential;
  %%%%%%%%%%%%%%%%%%%%%%%%%%%% 
   \footnote{This superpotential is consistent with R-symmetry $U(1)_R$,
   in which $Z_{L,R}$, $X_{L,R}$, $Y_{L,R}$, $\phi_{L,R}$ and
$\sigma^{(i)}_{L,R}$
   have  $U(1)_R$ charges, 2, 2, 2, 0 and 0.}
 %%%%%%%%%%%%%%%%%%%%%%%%%%%%%%%
 \begin{eqnarray}
  W &=& Z_L (\phi^2_L - 3v_L^2) + Z_R (\phi^2_R - 3v_R^2) \nonumber \\
     && +  X_L ( a_{(i)} \phi_L \sigma_L^{(i)} \phi_L )  
	  +  X_R ( a'_{(i)}\phi_R \sigma_R^{(i)} \phi_R )  \nonumber \\
	  &&+  Y_L ( b_{(i)} \phi_L \sigma_L^{(i)}   \phi_L)  
	  +  Y_R ( b'_{(i)} \phi_R \sigma_R^{(i)} \phi_R)  \ .
  \label{Super}
 \end{eqnarray}	
\noindent
Here, the fields $Z_{L,R}$, $X_{L,R}$ and $Y_{L,R}$ are all singlets 
of $O(3)_L\times O(3)_R$.
The parameters $v_L$ and $v_R$ can be of order the gravitational scale
 in principle,
since there is no symmetry to suppress them.  We should, however,
assume that they are of order the flavor scale $M_f$ to obtain
${\cal O}(1)$ effective Yukawa couplings of Higgs $H$ and $\bar H$.

We obtain vacuum-expectation values from the superpotential eq.(\ref{Super}),
 \begin{equation}
  <\phi_L> \equiv \left ( \matrix{1\cr 1 \cr 1 \cr} \right ) v_L \ , \qquad
  <\phi_R> \equiv \left ( \matrix{1\cr 1 \cr 1 \cr} \right ) v_R \  .
  \label{vac}
 \end{equation}
 Notice that only with the first two terms in eq.(\ref{Super}) we have 
 $O(3)_L\times O(3)_R$ global symmetry and hence unwanted Nambu-Goldstone
multiplets appear
  in broken vacua.  The couplings to  the explicit breakings 
  $\sigma_{L,R}^{(i)}$ are necessary to eliminate the Nambu-Goldstone multiplets
  in the low energy spectrum, which determine  vacuum-expectation values 
  of $\phi_L$ and $\phi_R$ as in eq.(\ref{vac}).  Here, we should quote a
work by
  Barbieri et al. \cite{O32}, who have also proposed a similar
vacuum-misalignment
  mechanism.
  
   With the non-vanishing $<\phi_L>$ and  $<\phi_R>$
   in eq.(\ref{vac}), the Dirac masses of charged leptons
   arise from a superpotential,
   \begin{equation}
      W = \frac{\k_E}{M_f^2} (\bar e\phi_R) (\phi_L \ell)\bar H  .
   \end{equation}
   \noindent
   This produces so-called "democratic" mass matrix of the charged leptons,
   \begin{equation}
  \widehat m_E = \k_E \left ( \frac{v_L v_R}{M_f^2}\right )
    \left (\matrix{1 & 1 & 1 \cr
        1 & 1 & 1 \cr  1 & 1 & 1 \cr }  \right ) <\bar H> .
		\label{lmass0}
  \end{equation}
	Diagonalization of this mass matrix yields large lepton mixings
 as shown in ref.\cite{Demo2} and one non-vanishing eigenvalue, $m_\tau$.
	The masses of e and $\mu$ are derived from distortion of the 
	"democratic" form of mass matrix in eq.(\ref{lmass0}), which is
	given by a superpotential containing the explicit 
	$O(3)_L\times O(3)_R$ breaking parameters  $\sigma_{L,R}^{(i)}$,
%%%%%%%%%%%%%%%%%%%%%%%%%%%%%%%%%%%
	\footnote{We may also have such terms as 
	$D^\ell_{ij}(\bar e \sigma^{(i)}_R \sigma^{(j)}_R \phi_R) (\phi_L \ell)$ or
	$E^\ell_{ij}(\bar e \phi_R) (\phi_L\sigma^{(i)}_L \sigma^{(j)}_L \ell)$
	 in the curry bracket of eq.(\ref{dW}). These terms and higher terms of
$\sigma$
 fields can be, however, absorbed in terms in eq.(\ref{dW}).
 In this process, the coupling parameters $A_i^\ell$, $B_i^\ell$ and
$C_{ij}^\ell$
	remain of ${\cal O}(1)$.}
%%%%%%%%%%%%%%%%%%%%%%%%%%%%%%%%%%%
 \begin{equation}
  \delta W = \frac{\k_E}{M_f^2} \left \{ A_i^\ell (\bar e
\sigma^{(i)}_R\phi_R) (\phi_L \ell)
   + B_i^\ell (\bar e \phi_R) (\phi_L \sigma^{(i)}_L \ell)
   + C_{ij}^\ell (\bar e \sigma^{(i)}_R \phi_R) (\phi_L \sigma^{(j)}_L
\ell)\right \}\bar H  .
   \label{dW}
   \end{equation}
   \noindent
   Then, the charged lepton mass matrix is given by
   
  \begin{eqnarray}
   &&\widehat m_E = \k_E \left ( \frac{v_L v_R}{M_f^2}\right )  <\bar H>
   \Biggl\{ \left (\matrix{1 & 1 & 1 \cr  1 & 1 & 1 \cr  1 &  1 &  1 \cr 
}\right )
  +  A_1^\ell \left (\matrix{1 & 1 & 1 \cr  1 & 1 & 1 \cr -2 & -2 & -2 \cr
} \right ) \delta_R 
  +  B_1^\ell \left (\matrix{1 & 1 & -2 \cr  1 & 1 & -2 \cr 1 & 1 & -2 \cr
} \right ) \delta_L 
	          \nonumber \\  \nonumber \\
 &&+ C_{11}^\ell\left (\matrix{1 & 1 & -2 \cr 1 & 1 & -2 \cr -2 & -2 & 4
\cr } \right )\delta_L\delta_R
+  A_2^\ell \left  (\matrix{1 & 1 & 1 \cr  -1 & -1 & -1 \cr 0 & 0 & 0 \cr 
} \right ) \e_R  
+  B_2^\ell \left  (\matrix{1 & -1 & 0 \cr  1 & -1 & 0 \cr 1 & -1 & 0 \cr }
\right ) \e_L 
	   \nonumber \\ \nonumber \\
&&+C_{12}^\ell \left (\matrix{1 & -1 & 0 \cr 1 & -1 & 0 \cr -2 & 2 & 0 \cr
} \right )\delta_R\e_L
+ C_{21}^\ell \left (\matrix{1 & 1 & -2 \cr -1 & -1 & 2 \cr 0 & 0 & 0 \cr }
\right )\delta_L\e_R
  + C_{22}^\ell \left (\matrix{1 & -1 & 0 \cr -1 & 1 & 0 \cr 0 & 0 & 0 \cr
} \right )\e_L\e_R	 
	\Biggr\} .
	\label{lmass}
  \end{eqnarray}
  
  \noindent
 The mass eigenvalues of this lepton mass matrix are
 \begin{equation}
  m_\tau \simeq 3\k_E \frac{v_L}{M_f}\frac{v_R}{M_f}<\bar H> , \qquad
  \frac{m_\m}{m_\tau} \simeq {\cal O}(\d_L \d_R), \qquad 
  \frac{m_e}{m_\tau} \simeq {\cal O}(\e_L \e_R),
 \end{equation}
\noindent where we assume that all coupling parameters 
$A_i^\ell$, $B_i^\ell$ and $C_{ij}^\ell(i,j=1,2)$ are of ${\cal O}(1)$.
Since the analysis on the quark mass matrices below shows  preferable values
 $\d_R\simeq 1$ and $\e_R\simeq 0.1-1$, we take 
 $\d_L\simeq 0.1$ and $\e_L\simeq 10^{-3}-10^{-2}$.
 
  We have an additional contribution to the neutrino mass matrix in
eq.(\ref{nuemass0}) as
   \begin{equation}
  \delta W = \frac{H^2}{M}\ell\left ( \b \frac{\phi_L}{M_f}
\frac{\phi_L}{M_f}\right )\ell .
   \end{equation}
   \noindent
The neutrino mass matrix is now given by 
\begin{eqnarray}
  \widehat m_\n &=&
  \frac{<H>^2}{M}\Biggl \{ \left (\matrix{1 & 0 & 0 \cr
        0 & 1 & 0 \cr  0 & 0 & 1 \cr  }  \right ) + 
		 \a_{(1)} \left (\matrix{1 & 0 & 0 \cr
        0 & 1 & 0 \cr  0 & 0 & -2 \cr  }   \right ) \delta_L +
		\a_{(2)} \left (\matrix{1 & 0 & 0 \cr
        0 & -1 & 0 \cr  0 & 0 & 0 \cr  } \right ) \e_L \nonumber \\
		& & +\b \left (\frac{v_L}{M_f}\right )^2 
		\left (\matrix{1 & 1 & 1 \cr  1 & 1 & 1 \cr  1 &  1 &  1 \cr  }\right )
		 \Biggr \} .
	\label{nuemass}
 \end{eqnarray}
 
 To see the  neutrino mixing, we take 
  the hierarchical base by applying an orthogonal transformation:  
   the charged lepton and neutrino  mass matrices are written as
%%%%%%%%%%%%%
\begin{equation}
  F^T \widehat m_E  F = 
  \k_E \left ( \frac{v_L v_R}{M_f^2}\right )  <\bar H>
  \left ( \matrix{2 C_{22}^\ell\e_L \e_R & 2\sqrt{3} C_{21}^\ell\e_R\d_L  
                          & \sqrt{6} A_2^\ell \e_R \cr
						  &        &        \cr
	2\sqrt{3}  C_{12}^\ell \e_L \d_R & 6 C_{11}^\ell \d_L \d_R  & 3\sqrt{2}
A_1^\ell \d_R  \cr 
	                      &            &             \cr
	 \sqrt{6} B_2^\ell \e_L  & 3\sqrt{2} B_1^\ell\d_L  & 3  \cr  
	 } \right )_{RL}  ,
\label{lmassh}
\end{equation}
 \noindent and 
\begin{equation}
     F^T \widehat m_\n  F =  \frac{<H>^2}{M}
	 \left ( \matrix{ 1+\a_{(1)}\d_L & \frac{1}{\sqrt{3}} \a_{(2)}\e_L  & 
	     \sqrt{\frac{2}{3}} \a_{(2)}\e_L \cr
		  &        &        \cr
	 \frac{1}{\sqrt{3}} \a_{(2)}\e_L & 1- \a_{(1)}\d_L  & \sqrt{2}
\a_{(1)}\d_L  \cr 
	      &        &        \cr
	 \sqrt{\frac{2}{3}} \a_{(2)}\e_L & 	\sqrt{2} \a_{(1)}\d_L  & 1 + 
	 3 \b \left (\frac{v_L}{M_f}\right )^2  \cr  
	 } \right )_{LL}  , 
	 \label{nuemassh}
\end{equation}	
\noindent where
\begin{equation}
 F= \left (\matrix{{1\o \sqrt{2}} &  {1\o \sqrt{6}} & {1\o \sqrt{3}} \cr
              -{1\o \sqrt{2}} &  {1\o \sqrt{6}} & {1\o \sqrt{3}} \cr
                      0       & -{2\o \sqrt{6}} & {1\o \sqrt{3}} \cr
                                         } \right )  .
  \end{equation}
  
\noindent
 The large neutrino mixing angle between  $\n_\m$ and $\n_\tau$
 indicated from the atmospheric neutrino oscillation is obtained \cite{Demo2}
 %%%%%%%%%%%%%%%%%%
 \footnote{The mixing matrix $U$ which determines neutrino oscillations is
defined as
  $U=U_\ell^\dagger U_\n$, where $U_\ell$ and $U_\n$ are diagonalization
matrices for
  the charged lepton 
  and neutrino mass matrices in eq.(\ref{lmassh}) and eq.(\ref{nuemassh}),
respectively
  (see ref.\cite{Demo2} for the definition of $U_\ell$ and $U_\n$). }
%%%%%%%%%%%%%%%%%%%%%%%%
if 
\begin{equation}
 \b \left (  \frac{v_L}{M_f}\right )^2 \ll \a_{(1)} \d_L  \ .
 \label{cond}
\end{equation}

  We also see  large 
  neutrino mixings between $\n_e$ and $\n_{\m, \tau}$
 from  the mass matrices eq.(\ref{lmassh}) and  eq.(\ref{nuemassh})
 for $\b (v_L/M_f)^2\leq \a_{(2)}\e_L$.
  %%%%%%%%%%%%%%%%%%%%%%%%%%%%%%%%%%
 By using 
 $\Delta m^2_{23}(\equiv m_{\n_3}^2-m_{\n_2}^2) \simeq 10^{-3} \eV^2$  for
the $\n_\mu-\n_\tau$
 oscillation \cite{SKam} (which corresponds to $m_{\n_i}={\cal O}(0.1) \eV$
 %%%%%%%%%%%%%%%%%%%%%%%%%%%%%%%%%%%%%%%%%
 \footnote{The constraint of $m_{\n_i}$ is given by the new upper limit
  of the  double $\b$ decay experiment $m^{ee}_{eff}< 0.2\eV$ \cite{Double}.
  The present model is viable without any accidental cancellation \cite{Ellis}.}
 %%%%%%%%%%%%%%%%%%%%%%%%%%%%%%%%%%%%%%%%%
 ), 
 $\d_L\simeq 0.1$ and  $\e_L\simeq 10^{-3}-10^{-2}$, we obtain
 %%%%%%%%%%%%%%%%%%%%%%%%%%%%%%%%%%%%%%%%%%%%%%%%%%%
 \footnote{If we use $\e_L/\d_L\simeq 0.1$ suggested from quark mixings, we get
 $\Delta m^2_{12}\simeq 10^{-4} \eV^2$. We consider that this is still
consistent with
 the large MSW solution taking account of ${\cal O}(1)$ ambiguity.}
 %%%%%%%%%%%%%%%%%%%%%%%%%%%%%%%%%%%%%%%%%%%%%%%%%%%
 \begin{equation}
 \Delta m^2_{12} \simeq \frac{\e_L}{\d_L}\Delta m^2_{23} \simeq
10^{-5}-10^{-4} \eV^2 ,
 \end{equation}
 \noindent for the $\n_e-\n_{\mu,\tau}$ oscillation.
 This  is consistent with the large angle MSW solution \cite{MSW}
 to the solar neutrino problem.
%%%%%%%%%%%%%%%%%%%%%%%%%%%%%%%%%%%%%%%%%%%%%%%%%%
\footnote{The current data \cite{NewSKam} of Super-Kamiokande experiments
give $\Delta m^2_{12} \simeq 2\times 10^{-5}-2\times 10^{-4} \eV^2$ and
 $\sin^2 2\theta_{12}=0.60-0.97$ at the $99\%$ confidence level, for the
large MSW 
 solution.}
%%%%%%%%%%%%%%%%%%%%%%%%%%%%%%%%%%%%%%%%%%%%%%%%%%
 For $\a_{(2)}\e_L \ll \b (v_L/M_f)^2 \ll a_{(1)}\d_L $ we have small
mixings between
  $\n_e$ and $\n_{\m,\tau}$.
  However, we obtain, in this case, $\Delta m^2_{12}\gg
10^{-5}-10^{-4}\eV^2$ which is too large
  for the small angle MSW solution \cite{MSW}.
  Therefore, we consider that this is an unlikely case.
  From the above argument we may conculde that $(v_L/M_f)\leq
\sqrt{\e_L}\simeq 0.03-0.1$.
  This implies small  $\tan\b=<H>/<\bar H>\simeq {\cal O}(1)$ unless
$\k_E(v_R/M_f)$
  is very large.
  %%%%%%%%%%%%%%%%%%%%%%%%%%%%%%%

 %%%%%%%%%%%%%%%%%%%%%%%%%%%%%%%%%%%%%%%%%
 %%%%%%%%%%%%%% Quark Sector %%%%%%%%%%%%%
 %%%%%%%%%%%%%%%%%%%%%%%%%%%%%%%%%%%%%%%%%
 We now turn to the quark sector, in which three doublet quarks
 $q_i(i=1-3)$ transform as an $O(3)_L$ triplet while three down quarks
$\bar d_i(i=1-3)$
 and the three up quarks $\bar u_i(i=1-3)$ as $O(3)_R$ triplets. 
 The  mass matrix of  down quarks  is given in the hierarchical base by
 \begin{equation}
  F^T \widehat m_D  F = 
  \k_D \left ( \frac{v_L v_R}{M_f^2}\right )  <\bar H>
  \left ( \matrix{2 C_{22}^D\e_L \e_R & 2\sqrt{3} C_{21}^D\e_R\d_L  &
\sqrt{6} A_2^D \e_R \cr
       &        &         \cr 
	2\sqrt{3}  C_{12}^D \e_L \d_R & 6 C_{11}^D \d_L \d_R  & 3\sqrt{2} A_1^D
\d_R  \cr 
	   &        &         \cr 
	 \sqrt{6} B_2^D \e_L  & 3\sqrt{2} B_1^D\d_L  & 3  \cr  
	 } \right )_{RL}  . 
\label{Dmassh}
\end{equation}

\noindent
For up quarks,  the mass matrix is given by replacing
 $D$ and  $<\bar H>$ with $U$ and $<H>$, respectively, in eq.(\ref{Dmassh}).
 Quark mass eigenvalues are given by
 \begin{equation}
  \left (\frac{m^Q_2}{m^Q_3}\right )\simeq 2 (A_1^Q B_1^Q-C_{11}^Q) 
    \d_L \d_R \equiv X_2^Q \d_L \d_R,
  \label{mQ2}
 \end{equation}
 \noindent and 
\begin{eqnarray}
  \left (\frac{m^Q_1}{m^Q_3}\right )=&& {2\o 3}(A_2^Q B_2^Q C_{11}^Q+A_1^Q
B_1^Q C_{22}^Q 
  -C_{11}^Q C_{22}^Q - A_2^Q B_1^Q C_{12}^Q-A_1^Q B_2^Q C_{21}^Q +C_{12}^Q
C_{21}^Q) 
       \e_L \e_R  \nonumber \\
	   &&\equiv X_1^Q \e_L \e_R,
  \label{mQ1}
 \end{eqnarray}
 \noindent
 where $Q=D$ or $U$. 
 We assume  that  the parameters $A_i^Q$, $B_i^Q$, $C_{ij}^Q$ are of ${\cal
O}(1)$
 as in the case of the lepton sector.
  We must, however, take  
  $\k_U(v_R/M_f)  \simeq {\cal O}(10)$ for $(v_L/M_f)\simeq 0.03$ 
to obtain  the large mass of the top quark.

 The CKM mixing angles are given by
  \begin{eqnarray}
  &&\left |V_{us} \right | \simeq s_{12}^D - s_{12}^U  \ ,\nonumber \\
  &&\left |V_{cb} \right | \simeq s_{23}^D - s_{23}^U  \ ,\nonumber \\
  &&\left |V_{ub} \right | \simeq s_{13}^D - s_{12}^U s_{23}^D+s_{12}^U
s_{23}^U - s_{13}^U\ ,
  \end{eqnarray}
 where $s_{ij}^Q$ denotes $\sin \theta_{ij}^Q$ and
    \begin{eqnarray}
    &&s_{12}^Q \simeq  \frac{1}{\sqrt{3}}
	     \frac{B_1^Q B_2^Q+2 C_{11}^Q
C_{12}^Q\d_R^2}{(B_1^Q)^2+2(C_{11}^Q)^2\d_R^2}
		 \ \frac{\e_L}{\d_L} \equiv \frac{1}{\sqrt{3}} Y_{12}^Q \
\frac{\e_L}{\d_L} \ , \nonumber \\
	&&s_{23}^Q \simeq  
	     \frac{\sqrt{2} B_1^Q}{1 + 2 (A_1^Q)^2 \d_R^2} \ \d_L 
		  \equiv \sqrt{2} Y_{23}^Q  \ \d_L \ , \nonumber \\
    &&s_{13}^Q \simeq  
	     -\sqrt{{2\o 3}}\frac{B_2^Q+ 2 A_1^Q C_{12}^Q\d_R^2}{1 + 2 (A_1^Q)^2
\d_R^2} \ \e_L 
		   \equiv  -\sqrt{{2\o 3}} Y_{13}^Q \ \e_L  \ . 
  \end{eqnarray}
  Putting the experimental quark mass ratios and CKM matrix elements:
  \begin{equation}
  \frac{m_d}{m_b}\simeq \l^4, \quad \frac{m_s}{m_b}\simeq \l^2, 
   \quad \left |V_{us} \right | \simeq \l,  \quad 
                                  \left |V_{cb} \right | \simeq \l^2 ,
 \end{equation}
 we obtain the order of parameters as follows:
 \begin{equation}
  \d_L \simeq \l^2, \qquad \d_R\simeq 1 \ , \qquad
  \e_L \simeq \l^3, \qquad \e_R \simeq \l \  ,
  \label{mag}
 \end{equation}
 \noindent  with $\l \simeq 0.2$.
 Here, we have assumed that
   $X_1^D$, $X_2^D$, $Y_{12}^D$, $Y_{23}^D$, $Y_{12}^U$ and $Y_{23}^U$ are
of ${\cal O}(1)$.
%%%%%%%%%%%%%%%%%%%%%%%%%%%%%%%%%%%%
 \footnote{To obtain the larger mass hierarchy 
 $m_u/m_t\simeq\l^8$ and $m_c/m_t\simeq\l^4$ in the up-quark sector 
   we must assume cancellation among  ${\cal O}(1)$ parameters in 
   $X_1^U$, $X_2^U$. A way of avoiding this fine-tuning is to introduce a
new $O(3)_R'$,
   under which up quarks $\bar u_i$ transform as a triplet.  In this case
   the up quark mass matrix depends on breaking parameters of the
additional $O(3)_R'$.}
%%%%%%%%%%%%%%%%%%%%%%%%%%%%%%%%%%%%%
 Then, we predict  $\left |V_{ub} \right |\simeq \e_L \simeq \l^3$,
  which is consistent with the experimental value \cite{PDG}. 
 The magnitudes of $\d_{L,R}$ and $\e_{L,R}$ in eq.(\ref{mag}) are what we
have taken
 in the discussion on the lepton sector.
    Thus our model is successful to explain  both lepton and quark mass
matrices.
	
%%%%%%%%%%%%%%%  Another Possiblity  %%%%%%%%%%%%
 We have considered, in this letter, a model where $\ell_i$ and $q_i$ belong to
 triplets of one $O(3)$ and $\bar e_i$, $\bar d_i$ and  $\bar u_i$ belong to
 triplets of the other  $O(3)$.  We should note here that  there is another
 interesting assignment that $\ell_i$ and $\bar d_i$ are triplets of the 
 $O(3)_L$ while $\bar e_i$, $q_i$ and  $\bar u_i$ transform as triplets of
the $O(3)_R$.
 At a first glance this model does not seem to work well,
 since the up quarks have $O(3)_R$-invariant degenerate masses as the neutrinos.
 However, this problem may be easily solved by imposing a discrete symmetry
 such as $Z_6$.  The $Z_6$ charges of relevant fields are shown in Table 1
 together with $O(3)_L\times O(3)_R$ representations. 
 With this $Z_6$ we obtain the "democratic" mass matrices for
 charged leptons, down quarks and up quarks, and the neutrinos have almost
degenerate
 mass as in the previous model.
 Remarkable point in this model is that the mass hierarchy in the up quark
sector is
 explained by taking a hierarchy
 $\d_R\simeq 0.1$ and $\e_R\simeq 0.01$.  The milder mass hierarchies in
 charged lepton and down quark sectors are obtained by taking
 $\d_L\simeq 1$ and $\e_L\simeq 0.1$. The CKM matrix is determined
 by   $O(3)_R$ breaking parameters $\d_R$ and $\e_R$, which  turns out to
be also
 consistent with the observations taking account of ${\cal O}(1)$ ambiguity.
 As for the neutrino mass matrix we obtain almost the same as before,
 except that the masses $m_{\n_i}$ are reduced to $m_{\n_i}\simeq {\cal
O}(0.03) \eV$
 due to the larger value of $\d_L$.
 %%%%%%%%%%%%%%%%%%%%%%%%%%%%%%%%%%%%%%%
%%%%%%%%%%%%%%%%%%%%%%%%%%%%%%%%%%%%%%%
%%%%%%%%%%%%%%%  Summary  %%%%%%%%%%%%
 
In this letter we have assumed specific forms of  
       $\Sigma^{(i)}_{L,R}$ in eq.(\ref{S1}) and eq.(\ref{S2}).
  However,  
   even if $\Sigma^{(i)}_{L,R}$  have general forms,
   $(\Sigma^{(1)}_{L,R})_{ij}=(a)_{ij}\d_{L,R}$ and 
   $(\Sigma^{(2)}_{L,R})_{ij}=(b)_{ij}\e_{L,R}$, our main conclusion does
not change 
   as long as $\e_L,\ \d_L \ll 1$ and $\e_R,\ \d_R \leq 1$.
   With the general form of $\Sigma^{(i)}_{L,R}$ we have vacuum-expectation
values of 
   $\phi_L$ and $\phi_R$ as 
   \begin{equation}
  <\phi_L> = \left ( \matrix{a\cr b \cr c \cr} \right ) v_L \ , \qquad
  <\phi_R> = \left ( \matrix{a'\cr b' \cr c' \cr} \right ) v_R \ ,
  \label{vacgene}
 \end{equation}
 \noindent with $a^2+b^2+c^2=a'^2+b'^2+c'^2=3$.
 First of all, mass hierarchies in the charged lepton and quark sectors, and 
 the small CKM mixing angles are explained by the hierarchy 
 $\e_L\ll \d_L \ll 1$.
 On the other hand, we see that neutrino mixings are very large as long as
  vacuum-expectation values
 $a$, $b$ and $c$ in eq.(\ref{vacgene}) are of ${\cal O}(1)$.
 Therfore, the large neutrino mixings are generic predictions in the present
 mechanism of $O(3)_L\times O(3)_R$ breaking.
 
  Finally, we should note that $\Sigma^{(i)}_{L,R}$ can be regarded 
  as vacuum-expectation values of dynamical $\Sigma^{(i)}_{L,R}$ fields
  inducing the spontaneous  $O(3)_L\times O(3)_R$ breaking.
  In this case we have six massless Nambu-Goldstone multiplets.
  However, the breaking scale $F_{L,R}$ are stringently constrained from
various 
  experimental results as $F_{L,R}>10^9 \G$ \cite{FY}. On the contrary,
   the flavor scale $M_f$ may be as low as $10 {\rm TeV}$  in the case of
explicit breaking.

 \vskip 10 mm
\noindent
{\large \bf Acknowledgements}

 MT and TY are supported in part by the Grants-in-Aid of the 
 Ministry of Education of Japan (Nos.10640274, 7107).
%\newpage
%%%%%%%%%%%%%%%%%%%%%%%%%%%%%%%%%%%%%%%%%%%%%%%%%%

\newpage
%%%%%%%%%%%%%%%%%%%%%%%%%%%%%%%%%%%%%%%%%%%%
%%%%%%%%%%%%%%    Table 1   %%%%%%%%%%%%%%%%
%%%%%%%%%%%%%%%%%%%%%%%%%%%%%%%%%%%%%%%%%%%%
\begin{table}
\hskip 2  cm
\begin{tabular}{ | r| r| r| r| r| r| r| r| r| r| r|} \hline
        &    & &  &      &        &         &        &        &   &        
   \\
 & $\ell_i\ $ & $\bar e_i\ $ &  $q_i\ $  &  $\bar d_i\ $ & $\bar u_i\ $ &
  $\phi_L\ $ &  $\phi_R\ $   & $H, \ \bar H$ &  $\sigma_{L}^{(1),(2)}$ &
$\sigma_{R}^{(1),(2)}$ \\
  & & &       &        &        &          &        &     &   & \\ \hline
  & & &       &        &        &          &        &     &   &  \\
$O(3)_L$ &$\bf 3\ $ & $\bf 1\ $ &$\bf 1\ $ & $\bf 3\ $ &  $\bf 1\ $  
           & $\bf 3\ \ $ & $\bf 1\ \ $ & $\bf 1\quad $ & $\bf 5\quad$  &
$\bf 1\quad$ \\
  & & &     &      &        &          &        &      &      &    \\ 
 $O(3)_R$ & $\bf 1\ $ & $\bf 3\ $  & $\bf 3\ $ & $\bf 1\ $ &  $\bf 3\ $ 
          & $\bf 1\ \ $ & $\bf 3\ \ $ & $\bf 1\quad $ & $\bf 1\quad$ & $\bf
5\quad$\\
   & & &     &      &        &          &        &      &     &   \\ 
$Z_6\quad$ & $1\ $ & $1\ $ & $1\ $ & $1\ $ & $1\ $  & $1\ \ $ & $1\ \ $ &
$2\quad $ & $0\quad$ 
& $0\quad$\\
  &      &   &   &     &         &          &    &   &  &  \\      \hline
\end{tabular}
\caption{Representations of relevant fields in $O(3)_L\times O(3)_R$
and their $Z_6$ charges.  The operators $\bar u q H$ and $\ell(\phi_L
\phi_L)\ell H^2$
are suppressed by this $Z_6$ symmetry.}
\end{table}
%%%%%%%%%%%%%%%%%%


\vskip 1 cm
\begin{thebibliography}{1}
 \bibitem{Abe}
 C. D. Froggatt and H. B. Nielsen,  Nucl. Phys. {\bf B147} (1979) 277;\\
 J. Bijnens and C. Wetterich,  Nucl. Phys. B {\bf 292} (1987) 443.

 
 \bibitem{NonAbe}
  S. Pakvasa and H. Sugawara, Phys. Lett. {\bf B73}  (1978) 61;\\
  F. Wilczek and A. Zee, Phys. Rev. Lett. {\bf  42} (1979) 421;\\
  G. B. Gelmini, J. M.  Gerard, T.  Yanagida and G.  Zoupanos, 
        Phys. Lett. {\bf B135}  (1984) 103.
 
  \bibitem{O31} 
  C. D. Carone and M. Sher, Phys. Lett. {\bf B420}  (1998) 83;\\
  E. Ma,  hep-ph/9812344;\\
  C. Wetterich, hep-ph/9812426;\\
  Y-L. Wu,  hep-ph/9901320.

  
 \bibitem{O32}
 R. Barbieri, L. J. Hall, G. L. Kane and G. G. Ross, hep-ph/9901228.

\bibitem{SKam}
Super-Kamiokande Collaboration, Y. Fukuda et al, Phys. Rev. Lett. {\bf 81} 
 (1998) 1562; hep-ex/9812014.

\bibitem{Double}
   H.V. Klapdor-Kleingrothaus, 
   Invited talk at
   % in  Proceedings of 
   the 18th International
   Conference on Neutrino Physics and Astrophysics, Takayama, Japan,
  % ed.  ,  World Scientific, xxx 
   (1998);\\
   Heidelberg-Moscow Collaboration, L. Baudis et al., hep-ex/9902014.

\bibitem{Demo1}
  H. Harari, H. Haut and J. Weyers, Phys. Lett. {\bf 78} B (1978) 459;\\
  Y. Koide, Phys. Rev. D 28 (1983) 252;  D {\bf 39}  (1989) 1391.
 
  \bibitem{Demo2}
 M. Fukugita, M. Tanimoto and  T. Yanagida, Phys. Rev. D {\bf 57} (1998) 4429;
 \\  hep-ph/9809554, to be published in Phys. Rev. D;  hep-ph/9903484;\\
  See also H. Fritzsch and Z. Xing, Phys. Lett.  {\bf 372B} (1996) 265.
  
 \bibitem{Ellis} 
 J. Ellis and S. Lola, hep-ph/9904279.
  
 \bibitem{MSW} 
  L. Wolfenstein, Phys. Rev. {\bf  D17} (1978) 2369;\\
 S.P. Mikheyev and A.Yu. Smirnov, Yad. Fiz. {\bf 42} (1985) 1441.

\bibitem{NewSKam}
T. Kajita, Invited talk at the conference ``Beyond the Desert'',
Tegernsee, Germany, June 6-12, 1999.				
 
 
 \bibitem{PDG} 
  Particle Data Group, EPJ C{\bf 3}  (1998) 1.
  
  \bibitem{FY} 
   M. Fukugita and  T. Yanagida, Phys. Rev. Lett. {\bf  55} (1985) 2645.
   
\end{thebibliography}
\end{document}